\documentclass[12pt]{article}
\usepackage{amsfonts}
\usepackage{mathtools}
\usepackage{slashed}
\usepackage{amsmath,amssymb}
\usepackage{graphicx,color}

\newcommand{\be}{\begin{equation}}
\newcommand{\ee}{\end{equation}}
\newcommand{\bea}{\begin{eqnarray}}
\newcommand{\eea}{\end{eqnarray}}

\def\a {\alpha}
\def\b {\beta}
\def\g {\gamma}

\begin{document}
\hfill AEI-2014-012\\

\centerline{\bf \Large Schwinger-Dyson approach to Liouville Field Theory}

\vspace{5mm}
\centerline{\bf }
\vspace{5mm}
\centerline{\bf Parikshit Dutta}

\vspace{5mm}
\begin{center}
{\it  Max-Planck-Institut
f\"ur Gravitationsphysik (Albert-Einstein-Institut),\\
Am M\"uhlenberg 1, D-14476 Potsdam, Germany}
\end{center}
\vspace{35mm}

\begin{abstract}
\noindent
We discuss Liouville field theory in the framework of Schwinger-Dyson approach
and derive a functional equation for the three-point structure constant.
We argue the existence of a second Schwinger-Dyson equation
on the basis of the duality between the screening charge operators
and obtain a second functional equation for the structure constant.
We discuss the utility of the two functional equations to fix the structure constant uniquely.
\end{abstract}
%\maketitle

%%%%%%%%%%%%%%%%%%%%%%%%%%%%%%%%%%%%%%%%%%%%%%%%%%
\newpage

\subsection*{Introduction}
Liouville  field theory has been a subject of intensive study during the last three decades.
It was initiated by the work of Polyakov \cite{poly}, where the quantization
of non-critical strings was reduced to a 2d field theory with an exponential interaction.
This model is integrable classically
due to the conformal symmetry, and there were several attempts to quantize it exactly \cite{ct}-\cite{JW1}.
However, canonical quantization based on the classical integrability
proved very problematic and could not yield the full two and three point correlation functions.

In the 90's, Dorn-Otto \cite{dorn}
and the brothers  Zamolodchikov  \cite{zz}, independently,
proposed an exact expression for the three point function
and provided a check of its consistency.
The exact expression was constructed from the residues of the correlation function
at some discrete points, where the calculation of the residues is possible by the path integral \cite{G-L}
and the Dotsenko-Fateev techniques \cite{fd}.
Since a continuation from discrete values, in general, is not unique, the obtained result
was called the DOZZ proposal.
However, soon after that, Teschner suggested a derivation
of the DOZZ formula with the help of the degenerated vertex operators \cite{t}.
One intriguing point about the DOZZ formula is that it has a dual set of poles,
which can not be seen form the path integral.
Later on, it was proposed in \cite{pr} that the dual set of poles is related
to the presence of an extra exponential potential which has to be introduced in the renormalized action.
In \cite{pr} it has been also claimed that the dual pole structure can fix the three point function uniquely.
However, no rigorous arguments for the presence of the extra exponential term were given.
The DOZZ proposal passed many stringent tests (see for example \cite{t1}-\cite{JW})
and it is believed  that the formula is correct,
however, the duality is still not well understood.
Our aim in the present paper is to investigate the duality in more detail.

In the previous paper \cite{pmn} we examined  the DeWitt equation
for Liouville field theory. It is interesting to note that this Schwinger-Dyson type approach
provides the same relation for the three point function that was used in \cite{dorn} to check the proposal.
In the present paper we employ this approach to analyze the duality of the theory.

We start with a path integral formulation of Liouville field theory and
present the most important points of this approach.
We then derive the Schwinger-Dyson equation, which
helps to analyze the residues of the correlation functions and leads
to a functional equation for the structure constant.
After this, we introduce a dual Schwinger-Dyson equation
with another screening charge operator and obtain additional residues and
a second functional equations for the structure constant.
In the remaining part of the paper we provide a formal derivation of
two functional equations, starting from the bare action.

\subsection*{Definitions and notations}

Liouville field theory is described by the action
\begin{align}\label{Action}
S[\phi]=\frac{1}{4\pi}\int d^{2}x\sqrt g \left(g^{kl}\partial_k\phi\,\,\partial_l\phi+
Q R \phi+4\pi\mu e^{2b\phi}\right)~,
\end{align}
where $\phi$ is the Liouville field, $R$ is the scalar curvature for the background metric $g_{kl}$,
$\mu$ and $b$ are constants and $Q=b+\frac{1}{b}$.

The partition function
\begin{align}
Z[J]= \int [\textit{D}\phi]\,e^{-S[\phi]+\int d^{2}x\sqrt {g(x)}J(x)\phi(x)}
\end{align}
for a delta function type source
\begin{equation}\label{delta-source}
J(x)=\sum_i^n \frac{2\alpha_{i}}{\sqrt{ g(x)}}\,\delta^{2}(x-x_{i})
\end{equation}
provides the correlation functions of Liouville field theory
\be\label{correlation functions}
Z\Big[\sum_i^n \frac{2\alpha_{i}}{\sqrt{ g(x)}}\,\delta^{2}(x-x_{i})\Big]=
\int [\textit{D}\phi]\,e^{-S[\phi]}\,\prod_{i=1}^{n}e^{2\alpha_{i}\phi(x_{i})}\equiv
\Big\langle\prod_{i=1}^{n}e^{2\alpha_{i}\phi(x_{i})}\Big\rangle~.
\ee

The Liouville field exponential $e^{2\alpha\phi(x)}$ is a primary field with conformal weight
\begin{equation}\label{conformal weight}
\Delta_\alpha=\alpha(Q-\alpha)~,
\end{equation}
and the three point correlation function has the standard CFT form
\begin{equation}\label{3-point function}
\Big\langle e^{2\alpha_1\phi(x_1)}\,e^{2\alpha_2\phi(x_2)}\,e^{2\alpha_3\phi(x_3)}\,\Big\rangle=
|x_{12}|^{2\g_3}\,
|x_{23}|^{2\g_1}\,|x_{31}|^{2\g_2}\,C(\alpha_1,\alpha_2,\alpha_3)~,
\end{equation}
where $\g_1=\Delta_{\a_1}-\Delta_{\a_2}-\Delta_{\a_3}$, $\g_2=\Delta_{\a_2}-\Delta_{\a_3}-\Delta_{\a_1}$,
$\g_3=\Delta_{\a_3}-\Delta_{\a_1}-\Delta_{\a_2}$  and $x_{ij}$ ($ij=12,\,23,\,31$)
are the distances between the points $x_i$ and $x_j$.
Below we assume that the Liouville filed is given on a sphere.

The Laplace-Beltrami operator on a sphere has only
one zero mode, namely a constant function $\phi_0$.
By splitting the Liouville field $\phi(x)$ into the
zero mode and its orthogonal complement $\phi(x)=\phi_0+\tilde\phi(x)\,$, with $\int d^2x \,\sqrt{g}\,\,\tilde\phi=0,\,$
one gets $\int d^2x \,\sqrt{g}\,R\,\phi =8\pi \phi_0$.

The integration over the zero mode $\phi_0$ in \eqref{correlation functions} is given by
\bea\label{zero mode integration}
\int d\phi_0 \,\,e^{2(\tilde\a-Q)\phi_0-U[\tilde\phi]\,e^{2b\phi_0}}=\Gamma\left(\frac{\tilde\a-Q}{b}\right)\,\frac{1}{2b}\,
U[\tilde\phi]^{\frac{Q-\tilde\a}{b}}~, \\
\label{screening charge}
\mbox{with} \qquad \tilde\a \equiv\sum_{i=1}^n\a_i~, \qquad U_b[\tilde\phi]\equiv\mu \int d^{2}x \sqrt{g(x)}\,e^{2b\tilde\phi(x)}~,
\eea
leads to \cite{G-L}
\be\label{after zero mode integration}
\Big\langle\prod_{i=1}^{n}e^{2\alpha_{i}\phi(x_{i})}\Big\rangle=\Gamma\left(\frac{\tilde\a-Q}{b}\right)\frac{1}{2b}\,
\int[D\tilde\phi]\,\,e^{-S_0[\tilde\phi]}\,\prod_{i=1}^n e^{2\a_i\tilde\phi(x_i)}\,
\left(U_b[\tilde\phi]\right)^{\frac{Q-\tilde\a}{b}}~,
\ee
where, $S_0[\tilde\phi]$ is the free-field action
\be\label{FF Action}
S_0[\tilde\phi]=\frac{1}{4\pi}\int d^{2}x\sqrt g \,\,g^{kl}\partial_k\tilde\phi\,\,\partial_l\tilde\phi~.
\ee

When $(Q-\tilde{\alpha})=mb$, with $m$ being a positive integer, one can perform the free-field path integration
in \eqref{after zero mode integration} and
reduce this calculation to the Dotsenko-Fateev integrals \cite{fd}, which for $n=3$ yields
\be\label{F-D integrals-1}
\int[D\tilde\phi]\,\,
e^{-S_0[\tilde\phi]}\,\prod_{i=1}^3 e^{2\a_i\tilde\phi(x_i)}\,\left(U_b[\tilde\phi]\right)^m=
|x_{12}|^{2\g_3}\,|x_{23}|^{2\g_1}\,|x_{31}|^{2\g_2}\,(-1)^m\,m!\,I_{m}(\a_1, \a_2, \a_3)~,
\ee
where $\g_i$ ($i=1,2,3$) are the same as in \eqref{3-point function} and
\begin{equation}\label{F-D Integrals}
I_{m}(\a_1, \a_2, \a_3)=\bigg(\frac{-\pi\mu}{\gamma(-b^{2})}\bigg)^{m}\frac{\prod_{j=1}^{m}\gamma(-jb^{2})}
{\prod_{k=0}^{m-1}[\gamma(2\alpha_{1}b+kb^{2}) \gamma(2\alpha_{2}b+kb^{2}) \gamma(2\alpha_{3}b+kb^{2})]}~,
\end{equation}
with $\gamma(u)=\Gamma(u)/\Gamma(1-u)$.
Due to the poles of  the gamma function in \eqref{after zero mode integration} at $(Q-\tilde{\alpha})=mb$,
equation \eqref{F-D Integrals} defines the residues of the structure constants
$C(\alpha_1,\alpha_2,\alpha_3)$.
Using the form of the residues, the authors of \cite{dorn} and \cite{zz} were able to
construct $C(\alpha_1,\alpha_2,\alpha_3)$ as a meromorphic function
\begin{eqnarray}\label{DOZZ}
&& \!\!\!\!\!\! \!\!\!\!\!\! \!\!\!\!\!\!\!\!\!\!\!\! \!\!\!\!\!\!\!\!\!\!\!\!
C(\alpha_{1},\alpha_{2},\alpha_{3})=\bigg[\pi\mu\gamma(b^{2})b^{2-2b^{2}}\bigg]^{(Q-\tilde{\alpha})/b}\times\notag\\
&& \!\!\!\!\!\! \!\!\!\!\!\! \!\!\!\!\!\!\!\!\!\!\!\! \!\!\!\!\!\!\!\!\!\!\!\!
\frac{\Upsilon'(0)\,\Upsilon(2\alpha_{1})\Upsilon(2\alpha_{2})\Upsilon(2\alpha_{3})}{\Upsilon(\alpha_{1}+\alpha_{2}+
\alpha_{3}-Q)\Upsilon(\alpha_{1}+\alpha_{2}-\alpha_{3})\Upsilon(\alpha_{2}+\alpha_{3}-\alpha_{1})\Upsilon(\alpha_{3}+
\alpha_{1}-\alpha_{2})}~,
\end{eqnarray}
where  $\Upsilon(x)$ is given by the integral representation
\be\label{Upsilon}
\Upsilon(x)=\exp\left[\int_0^\infty \frac{dt}{t}\,\left(\left(\frac{Q}{2}-x\right)^2\,e^{-t}-
\frac{\sinh^2\left(\frac{Q}{2}-x\right)\frac{t}{2}}{\sinh \frac{b t}{2}\,\,\sinh\frac{t}{2b}}\right)\right]~.
\ee

Since such kind of analytical continuation,
in general, is not unique, there were several other efforts to check the validity of \eqref{DOZZ}.

Our aim is to apply the Schwinger-Dyson equation and find new functional relations between the correlation functions,
which can fix the structure constant $C(\alpha_1,\alpha_2,\,\alpha_3)$ in an
alternative way.

\subsection*{First functional equation}

The translation invariance of the path integral measure
\begin{equation}
\int [\textit{D}\phi]\frac{\delta}{\delta\phi(x)}\,\,e^{-S[\phi]+\int d^{2}x\sqrt{ g(x)}J(x)\phi(x)}=0
\end{equation}
leads to the Schwinger-Dyson type equation
\begin{equation}
 \,J(x)\,Z[J]-\bigg\langle\frac{\delta S}{\delta\phi(x)}\bigg\rangle=0~,
\end{equation}
which for Liouville field theory takes the form
\begin{equation}\label{SD-eq}
 J(x)Z[J]=-\frac{1}{2\pi}\Delta\frac{\delta Z[J]}{\delta J(x)} +\frac{1}{4\pi}QR(x)Z[J] +2\mu b Z[J_{x,b}]~,
\end{equation}
where $\Delta=\frac{1}{\sqrt g}\,\partial_a\,g^{ab}\sqrt g\,\partial_b$ is the Laplace-Beltrami operator and
\begin{align}
 J(y)_{x,b}=J(y)+2b\frac{1}{\sqrt{g(x)}}\delta^{2}(y-x)~.
\end{align}

The integration of \eqref{SD-eq} provides the following functional relation for the partition function
\begin{equation}\label{integral relation}
 \int d^{2}x\sqrt{g(x)} \Big(J(x)-\frac{1}{4\pi}QR(x)\Big)Z[J]=2\mu b\int d^{2}x \sqrt{g(x)}\, Z[J_{x,b}]~.
\end{equation}

Note that this equation is simply obtained also from
\begin{equation}
\int [\textit{D}\phi]\frac{\delta}{\delta\phi_{0}}\,\,e^{-S[\phi]+\int d^{2}x\sqrt{ g(x)}J(x)\phi(x)}=0~,
\end{equation}
which is the translation invariance of the measure only with respect to the zero mode.

Inserting the delta function type source \eqref{delta-source} in \eqref{integral relation} and integrating the curvature term as above, we find
\begin{align}\label{eq. 1}
\left(\tilde\alpha-Q\right)\Big\langle\prod_{i=1}^{n}e^{2\alpha_{i}\phi(x_{i})}\Big\rangle=\mu b \int d^{2}x \sqrt{g(x)}\, \Big\langle \prod_{i=1}^{n}e^{2\alpha_{i}\phi(x_{i})}e^{2b\phi(x)}\Big\rangle~.
\end{align}
This equation provides recursive relations between the correlations functions in an integral form.
Replacing $n$ by $n+1$ in \eqref{eq. 1} and setting one of $\alpha_i$'s equal to $b$, we get
\begin{align}
\left(\tilde\alpha+b-Q\right)\Big\langle\prod_{i=1}^{n}e^{2\alpha_{i}\phi(x_{i})}e^{2b\phi(x)}\Big\rangle=\mu b \int d^{2}y \sqrt{g(y)} \Big\langle \prod_{i=1}^{n}e^{2\alpha_{i}\phi(x_{i})}e^{2b\phi(x)}e^{2b\phi(y)}\Big\rangle~,
\end{align}
which together with \eqref{eq. 1} yields
\begin{eqnarray}
&& \!\!\!\!\!\! \!\!\!\!\!\! \!\!\!\!\!\!\!\!\!\!\!\! \!\!\!\!\!\!\!\!\!\!\!\!
\left(\tilde\alpha-Q\right)\left(\tilde\alpha+b-Q\right)\Big\langle\prod_{i=1}^{n}e^{2\alpha_{i}\phi(x_{i})}
\Big\rangle\notag\\
&& \!\!\!\!\!\! \!\!\!\!\!\! \!\!\!\!\!\!\!\!\!\!\!\! \!\!\!\!\!\!\!\!\!\!\!\!
\quad \quad \quad \quad \quad \quad=(\mu b)^{2}\int d^{2}x\sqrt{g(x)}  \int d^{2}y \sqrt{g(y)} \Big\langle \prod_{i=1}^{n}e^{2\alpha_{i}\phi(x_{i})}e^{2b\phi(x)}e^{2b\phi(y)}\Big\rangle~.
\end{eqnarray}
Repeating this procedure $m$-times, we obtain
\begin{eqnarray}\label{it1}
&& \!\!\!\!\!\! \!\!\!\!\!\! \!\!\!\!\!\!\!\!\!\!\!\! \!\!\!\!\!\!\!\!\!\!\!\!
\prod_{j=0}^{m}(\tilde{\alpha}+j b-Q)\,\,\Big\langle\prod_{i=1}^{n}e^{2\alpha_{i}\phi(x_{i})}\Big\rangle\notag\\
&& \!\!\!\!\!\! \!\!\!\!\!\! \!\!\!\!\!\!\!\!\!\!\!\! \!\!\!\!\!\!\!\!\!\!\!\!
\quad \quad \quad \quad \quad\qquad\qquad =(b\mu )^{m+1}\idotsint\Big\langle\prod_{i=1}^{n}e^{2\alpha_{i}\phi(x_{i})}\prod_{j=0}^{m}e^{2b\phi(y_{j})}\Big\rangle d^{2}y_{j}\sqrt{g(y_{j})}~.
\end{eqnarray}

Form this equation follows that the $n$-point functions are singular at $\tilde{\alpha}=Q-mb$,
which can be seen from the zero mode integration \eqref{after zero mode integration} as well.
By \eqref{eq. 1} we also realize that the insertion of the screening charge operator
\begin{equation}\label{screening charge op}
U_b= \mu\int d^{2}x \sqrt{g(x)}\,e^{2b\phi(x)}~
\end{equation}
modifies the correlator by the $\tilde\a$ dependent constant factor.
Note that the conformal dimension of the operator \eqref{screening charge op} is equal to zero and its insertion does not change
the conformal properties of the correlation functions.

Let us consider equation \eqref{it1} for $n=3$ and also the same equation with the replacement $m$ by $m-1$.
The relation of the corresponding expressions yields
 \begin{eqnarray}\label{eq1}
&& \!\!\!\!\!\! \!\!\!\!\!\! \!\!\!\!\!\!\!\!\!\!\!\! \!\!\!\!\!\!\!\!\!\!\!\!
(\tilde{\alpha}-Q+mb)\idotsint\Big\langle\prod_{i=1}^{3}e^{2\alpha_{i}\phi(x_{i})}\prod_{j=0}^{m-1}e^{2b\phi(y_{j})}
\Big\rangle d^{2}y_{j}\sqrt{g(y_{j})}= \notag\\
&& \!\!\!\!\!\! \!\!\!\!\!\! \!\!\!\!\!\!\!\!\!\!\!\! \!\!\!\!\!\!\!\!\!\!\!\!
\quad \quad \quad \quad \quad \quad \quad \quad\mu b \idotsint
\Big\langle\prod_{i=1}^{3}e^{2\alpha_{i}\phi(x_{i})}\prod_{j=0}^{m}e^{2b\phi(y_{j})}\Big\rangle d^{2}y_{j}\sqrt{g(y_{j})}~,
\end{eqnarray}
and using again \eqref{it1}, we find
\begin{eqnarray}\label{eq2}
&& \!\!\!\!\!\! \!\!\!\!\!\! \!\!\!\!\!\!\!\!\!\!\!\! \!\!\!\!\!\!\!\!\!\!\!\!
\Big\langle\prod_{i=1}^{3}e^{2\alpha_{i}\phi(x_{i})}\Big\rangle\prod_{j=0}^{m}(\tilde{\alpha}+j b-Q)\notag\\
&& \!\!\!\!\!\! \!\!\!\!\!\! \!\!\!\!\!\!\!\!\!\!\!\! \!\!\!\!\!\!\!\!\!\!\!\!
=(b\mu )^{m}(\tilde{\alpha}-Q+mb)\idotsint\Big\langle\prod_{i=1}^{3}e^{2\alpha_{i}\phi(x_{i})}\prod_{j=0}^{m-1}e^{2b\phi(y_{j})}\Big\rangle d^{2}y_{j}\sqrt{g(y_{j})}~.
\end{eqnarray}
Similarly to \eqref{zero mode integration}, the zero mode integration in the right hand side of this equation provides
\begin{eqnarray}\label{eq3}
&& \!\!\!\!\!\! \!\!\!\!\!\! \!\!\!\!\!\!\!\!
\idotsint\Big\langle\prod_{i=1}^{3}e^{2\alpha_{i}\phi(x_{i})}\prod_{j=0}^{m-1}e^{2b\phi(y_{j})}\Big\rangle d^{2}y_{j}\sqrt{g(y_{j})}=
\Gamma\bigg(\frac{\tilde{\alpha}+m b-Q}{b}\bigg)\,\frac{1}{2b}\,\int [D\tilde{\phi}]\,\,e^{-S_{0}(\tilde{\phi})}\times \notag\\
&& \!\!\!\!\!\! \!\!\!\!\!\! \!\!\!\!\!\!\!\!\!\!\!\! \!\!\!\!\!\!\!\!\!\!\!\!
\quad \quad \quad \prod_{i=1}^{3}e^{2\alpha_{i}\tilde{\phi}(x_{i})}\bigg(\int d^{2}y\sqrt{g(y)} e^{2b\tilde{\phi}(y)}\bigg)^{m}\bigg(\mu \int d^{2}x\sqrt{g(x)}  e^{2b\tilde{\phi}(x)}\bigg)^{\frac{(Q-\tilde{\alpha}-m b)}{b}}~.
\end{eqnarray}

Now we introduce the parameter $\epsilon=\tilde{\alpha}-Q+mb$ and consider the limit $\epsilon \to 0$.
The left and right hand sides of \eqref{eq2} have the same coordinate dependent parts
defined by \eqref{3-point function} and \eqref{F-D integrals-1},
respectively. Canceling these parts on both sides of \eqref{eq2}, we find
\be\label{eq4}
\lim_{\epsilon\to 0}\big[\prod_{j=0}^m(\epsilon-j b) \, C(\alpha_{1},\alpha_{2},\alpha_{3})\big]
=\frac{1}{2}\,\,b^{m}(m!)(-1)^{m}I_{m}(\alpha_{1},\alpha_{2},\alpha_{3})~,
\ee
which simplifies to
\be\label{eq5}
\lim_{\epsilon\to 0}\big[\epsilon \, C(\alpha_{1},\alpha_{2},\alpha_{3})\big]
=\frac{1}{2}\,\,I_{m}(\alpha_{1},\alpha_{2},\alpha_{3})~.
\ee
Replacing the parameters here by $\a_1\mapsto\alpha_1+b$ and $m\mapsto m-1$,
we obtain
\be\label{eq5'}
\lim_{\epsilon\to 0}\big[\epsilon \, C(\alpha_{1}+b,\alpha_{2},\alpha_{3})\big]=
\frac{1}{2}\,\,I_{m-1}(\alpha_{1}+b,\alpha_{2},\alpha_{3})~,
\ee
and the ratio of \eqref{eq5'} and \eqref{eq5} yields the following equation
\begin{align}\label{s0}
\frac{C(\alpha_{1}+b,\alpha_{2},\alpha_{3})}{C(\alpha_{1},\alpha_{2},\alpha_{3})}=\frac{I_{m-1}(\alpha_{1}+
b,\alpha_{2},\alpha_{3})}{I_{m}(\alpha_{1},\alpha_{2},\alpha_{3})}~.
\end{align}
With the help of \eqref{F-D Integrals}, its explicit form becomes
\be\label{s1}
\frac{C(\alpha_{1}+b,\alpha_{2},\alpha_{3})}{C(\alpha_{1},\alpha_{2},\alpha_{3})}=
-\frac{\gamma(-b^{2})\gamma(b(2\alpha_{1}+b))\gamma(2b\alpha_{1})\gamma(b(\alpha_{2}+
\alpha_{3}-\alpha_{1}-b))}{\mu\gamma(b(\alpha_{1}+\alpha_{2}+\alpha_{3}-Q))\gamma(b(\alpha_{1}+\alpha_{2}-
\alpha_{3}))\gamma(b(\alpha_{1}+\alpha_{3}-\alpha_{2}))}~.
\ee

Though this functional equation for the structure constant was derived with the restriction $\tilde\a=Q-mb$, thus being a limitation of this procedure, we will argue later that the restriction can be removed under certain justifiable assumptions. However
it was shown by Teschner \cite{t}, that \eqref{s1} is valid for the entire set of parameters $(\alpha_{1},\alpha_{2},\alpha_{3})$. We expect that this could possibly be done independently by expressing the ratios of the structure constants as functional integrals and performing some field redefinitions. The explicit proof is still a point we are working on. In the next section we will give some arguments to suggest that this equation derived using our method can be shown to be valid over the entire parameter space under some assumptions.

\subsection*{Second functional equation}

The equation for the unit conformal weight
\be\label{conformal weight 1}
\alpha(Q-\alpha)=1
\ee
has two solutions: $\a=b$ and $\a=1/b$. Hence, the operator
\begin{align}\label{screening chrage'}
U_{1/b}=\tilde\mu\int d^{2}x\sqrt{g(x)} e^{\frac{2}{b}\phi(x)}~,
\end{align}
with constant $\tilde\mu$, has zero conformal weight, like the operator \eqref{screening charge op}.
The insertion of the screening charge operator \eqref{screening chrage'}
does not change the conformal properties of the correlation functions, and
we use this property to suggest a second equation for the structure constant.

If we start with the relation analogous to \eqref{eq. 1}\footnote{We justify this relation in the next section.}
\begin{equation}\label{sc2}
(\tilde{\alpha}-Q)\Big\langle\prod_{i=1}^{n}e^{2\alpha_{i}\phi(x_{i})}\Big\rangle=\frac{\tilde{\mu}}{b}\int d^{2}x\sqrt{g(x)}\Big\langle\prod_{i=1}^{n}e^{2\alpha_{i}\phi(x_{i})}e^{\frac{2}{b}\phi(x)}\Big\rangle~,
\end{equation}
and perform the same iterative scheme as above, we find the equation similar to \eqref{it1}
\bea\label{eq6}
\prod_{j=0}^{m}\Big(\tilde{\alpha}+\frac{j}{b}-Q\Big)\Big\langle\prod_{i=1}^{n}e^{2\alpha_{i}\phi(x_{i})}\Big\rangle=
~~~~~~~~~~~~~~~~~~~~~~~~~~~~~~~~~~~~~~~~\\ \nonumber
~~~~~~~~~\idotsint\Big(\frac{\tilde{\mu}}{b}\Big)^{m+1}\Big\langle\prod_{i=1}^{n}e^{2\alpha_{i}\phi(x_{i})}
\prod_{j=0}^{m}e^{\frac{2}{b}\phi(y_{j})}\Big\rangle d^{2}y_{j}\sqrt{g(y_{j})}~.
\eea

The second functional equation, obtained at $\tilde{\alpha}=Q-m/b$, then takes the form \eqref{s0}
\begin{align}\label{feq2b}
\frac{C(\alpha_{1}+1/b,\alpha_{2},\alpha_{3})}{C(\alpha_{1},\alpha_{2},\alpha_{3})}=
\frac{\tilde{I}_{m-1}(\alpha_{1}+1/b,\alpha_{2},\alpha_{3})}{\tilde{I}_{m}(\alpha_{1},\alpha_{2},\alpha_{3})}~,
\end{align}
where $\tilde{I}_m$ is also defined by the free-field Coulomb gas integral
\begin{equation}\label{F-D Integrals'}
\int [D\tilde\phi] \,e^{-S_{0}[\tilde\phi]}\prod_{i=1}^{3}e^{2\alpha_{i}\phi(x_{i})}
\left(U_{1/b}[\tilde\phi]\right)^m=
|x_{12}|^{2\g_3}\,|x_{23}|^{2\g_1}\,|x_{31}|^{2\g_2}\,(-1)^m\,m!\,\tilde I_{m}(\a_1, \a_2, \a_3)~.
\end{equation}
Thus, $\tilde{I}_m$ is obtained from \eqref{F-D Integrals} by the replacements $b\mapsto 1/b$, $\mu\mapsto \tilde\mu$.
Equation  (\ref{feq2b}) then reduces to
\be\label{s2}
\frac{C(\alpha_{1}+1/b,\alpha_{2},\alpha_{3})}{C(\alpha_{1},\alpha_{2},\alpha_{3})}=
-\frac{\gamma(-b^{-2})\,\gamma(\frac{2\alpha_{1}+1/b}{b})
\gamma(2\frac{\alpha_{1}}{b})\gamma(\frac{\alpha_{2}+\alpha_{3}-\alpha_{1}-1/b}{b})}{\tilde\mu\,\gamma(\frac{\alpha_{1}+\alpha_{2}+
\alpha_{3}-Q}{b})\gamma(\frac{\alpha_{1}+\alpha_{2}-\alpha_{3}}{b})\gamma(\frac{\alpha_{1}+\alpha_{3}-\alpha_{2}}{b})}~.
\ee

Note again that equations (\ref{s1}) and (\ref{s2}) in our scheme were obtained at
$\tilde{\alpha}=Q-mb$ and $\tilde{\alpha}=Q-m/b $, respectively.
While equation (\ref{s1}) was derived in \cite{t} without any constraints on the parameters, equation (\ref{s2}) was sugested to exist from the point of view of duality between degenerate fields. This was because, certain coloumb gas integrals which would come up in the computation using crossing symmetry, does not satisfy the coloumb gas constraint and is not calculable in principle. This was also noted in \cite{N}. Hence the second functional equation in \cite{t} is suggestive and not derived explicitly. In our case we derive them over the space of constrained parameters suggesting the existence of dual screening charge equation (important for performing the actual Coloumb integral). We note that (\ref{s1}) and (\ref{s2}) are two functional equations valid on infinite set of hyper-planes in three dimensional space of complex numbers i.e. $C^{3}$, defined by equations: $\tilde{\alpha}=Q-mb$ and $\tilde{\alpha}=Q-n/b$, $m,n\geq 0$. Then the complete functional equations for (\ref{s1}) and (\ref{s2}) can only have two functions $f_{1}(\tilde{\alpha}-Q)$ and $f_{2}(\tilde{\alpha}-Q)$ as a prefactor of R.H.S.of equation (\ref{s1}) and (\ref{s2}). Now one must note that if there is any isolated simple pole in the three point function, i.e. at $\tilde{\alpha}-Q=c$, for some non zero constant $c$, then taking one of the $\alpha_{i}$s to zero (lets say $\alpha_{3}$) and keeping $\alpha_{1}+\alpha_{2}=c$, using either equation (\ref{s1}) or (\ref{s2}) it can be seen that taking the limit and assuming a smooth limit exists :
\begin{eqnarray}
\lim_{\alpha_{3}\to 0,\tilde{\alpha}-Q=c}\left(\tilde\alpha-Q\right)\Big\langle\prod_{i=1}^{3}e^{2\alpha_{i}\phi(x_{i})}\Big\rangle &=& \lim_{\alpha_{3}\to 0,\tilde{\alpha}-Q=c}\mu b \int d^{2}x \sqrt{g(x)}\, \Big\langle \prod_{i=1}^{3}e^{2\alpha_{i}\phi(x_{i})}e^{2b\phi(x)}\Big\rangle~.\notag\\
\Rightarrow c\langle\prod_{i=1}^{2}e^{2\alpha_{i}\phi(x_{i})}\Big\rangle &=& \mu b \int d^{2}x \sqrt{g(x)}\, \Big\langle \prod_{i=1}^{2}e^{2\alpha_{i}\phi(x_{i})}e^{2b\phi(x)}\Big\rangle~
\end{eqnarray}
 Since for first equation above, both L.H.S. and R.H.S are singular (simple pole at $\tilde{\alpha}-Q=c$), after taking the limit, (since $\tilde{\alpha}-Q$ is kept constant, and only $\alpha_{3}$ taken to zero, the limit can be assumed to be taken smoothly as the pole is in the variable $\tilde{\alpha}-Q$ and any possible combination of $\alpha$s can be chosen which satisfy $\tilde{\alpha}-Q=c$), we again have R.H.S. singular and so L.H.S. singular as $c$ is non zero. This means there is a two point function or some state with $\alpha_{1}=\alpha_{2}=c/2$ whose inner product is infinite (different from delta function normalizabe which gives $\alpha_{1}=\alpha_{2}$). This certainly can not be. Hence one can safely assume that the three point function has simple poles only at points of the parameter space $\tilde{\alpha}-Q=-bn-m/b$, i.e. only at integer multiples of the screening charges (maximal analyticity). Under this extra condition, the three point function for Liouville theory can only be the DOZZ formula (known part deduced from the residues with maximal analyticity i.e. with no. extra simple poles other than the the points mentioned above) times a periodic function $g(\tilde{\alpha}-Q)$, such that $g(-mb-n/b)=1$ for $m,n$ non-negative integers. It is well know that for irrational $b$ this function has to be a constant as the periods are incommensurable. On the other hand the functions $f_{1}$, $f_{2}$ is then $g(\tilde{\alpha}+b-Q)/g(\tilde{\alpha}-Q)$ and $g(\tilde{\alpha}+1/b-Q)/g(\tilde{\alpha}-Q)$ respectively, since the rest of the piece is satisfied by the DOZZ formula or equivalently by the residues. As pointed out for irrational $b$ these functions are also constants (equal to $1$), hence the difference equations (\ref{s1}) and (\ref{s2}) can be analytically continued over the entire parameter space. These two equations thus lead to the DOZZ formula \eqref{DOZZ} uniquely \cite{t}.

Indeed, from the properties of $\Upsilon$-function \eqref{Upsilon},
\be\label{properties of upsilon}
\Upsilon(x+b)=\g(bx)b^{1-2bx}\Upsilon(x)~, \qquad
\Upsilon(x+1/b)=\g(x/b)b^{2x/b-1}\Upsilon(x)~,
\ee
follows that $C(\a_1,\a_2,\a_3)$ defined by \eqref{DOZZ} satisfies both (\ref{s1}) and (\ref{s2}),
if the dual cosmological constants are related by
\be\label{dual constant}
\tilde{\mu}=(\pi\mu\gamma(b^{2}))^{b^{-2}}/(\pi\gamma(b^{-2}))~.
\ee
Let $D(\a_1,\a_2,\a_3)$ be another solution of the same equations with irrational $b$. The ratio
$D/C$ then will be a function with two incommensurable periods. It is well known
that such a function is constant, which leads to $D=C$.

In the next section  we discuss arguments leading to the relation \eqref{sc2} and also analyze
a relative scaling properties of $\tilde \mu$ and $\mu$, obtained directly from the path integral.
The later appears consistent with \eqref{dual constant}.

\subsection*{Source of the 2nd functional equation}

Let us consider equation \eqref{after zero mode integration}
with $\tilde{\alpha}=Q-mb-1/b$. In this case,
we can not calculate the free-field path integral on the right hand side of \eqref{after zero mode integration},
since the power of $U_b[\tilde \phi]$ is not integer.
An integer power of $U_b[\tilde \phi]$ is obtained by insertion of the second screening charge
operator, which formally yields
\bea\label{insertion 2}
&& \int d^{2}x\sqrt{g(x)} \langle\prod_{i=1}^{n}e^{2\alpha_{i}\phi(x_{i})}e^{\frac{2}{b}\phi(x)}\rangle=
\Gamma(-m)\frac{1}{2b}\int D[\tilde{\phi}]\,e^{-S_{0}[\tilde{\phi}]}\times ~~~~~~~~~\\
&&
\quad \quad \quad \prod_{i=1}^{n}e^{2\alpha_{i}\tilde{\phi}(x_{i})}\bigg(\int d^{2}y\sqrt{g(y)} e^{\frac{2}{b}\tilde{\phi}(y)}\bigg)\bigg(\mu \int d^{2}x\sqrt{g(x)} e^{2b\tilde{\phi}(x)}\bigg)^m~.
\eea
The argument of $\Gamma$-function here is a negative integer number and
the free-field path integral provides only the residue of the correlator.
This trick gives us a hint that the correlators have poles not only at $Q-\tilde\a =mb$, but also at $Q-\tilde\a =m/b$.

Our aim here is to investigate the relation between
$$\int d^{2}x\sqrt{ g(x)} \langle\prod_{i=1}^{n}e^{2\alpha_{i}\phi(x_{i})}e^{\frac{2}{b}\phi(x)}\rangle~
 \quad \mbox{and} \quad ~\langle\prod_{i=1}^{n}e^{2\alpha_{i}\phi(x_{i})}\rangle ~,$$
that was proposed in the previous section by \eqref{sc2}.

The Schwinger-Dyson equation
obtained from the bare action of Liouville field theory,
similarly to \eqref{integral relation}, yields
\begin{equation}
(\tilde\alpha-Q_{B})\langle\prod_{i=1}^{n}e^{2\alpha_{i}\phi(x_{i})}\rangle=\mu_{B} b \int d^{2}x \sqrt{g(x)}\langle \prod_{i=1}^{n}e^{2\alpha_{i}\phi(x_{i})}e^{2b\phi(x)}\rangle~,
\end{equation}
where the index $B$ stands for the bare parameters. Note that $Q_{B}=1/b$ and the classical
conformal dimension of $e^{2\a\phi(x)}$ is $\alpha Q_{B}$. Thus,
classically there is only one screening charge given by \eqref{screening charge op}.
Replacing the bare Schwinger-Dyson equation by a renormalized one, we find
\begin{equation}\label{rsd}
(\tilde\alpha-Q_{R})\langle\prod_{i=1}^{n}e^{2\alpha_{i}\phi(x_{i})}\rangle=\mu_{R}\, b\beta \int d^{2}x \sqrt{g(x)}\langle \prod_{i=1}^{n}e^{2\alpha_{i}\phi(x_{i})}e^{2b\beta\phi(x)}\rangle~,
\end{equation}
with renormalized parameters $Q_{R}$, $\mu_{R}$ and $\beta$.

Taking into account that both sides of eq. \eqref{rsd} should have the same conformal transformation properties,  one
gets the following condition on the renormalized parameters \cite{dh}
\begin{equation}
1-Q_{R}\beta b +\beta^{2}b^{2}=0~.
\end{equation}
%or
%\begin{equation}
%Q_{R}=\frac{1}{\beta b}+\beta b
%\end{equation}
By this equation there are two choices for $\b$, $\,\beta_1=1$ and $\,\b_2=1/b^{2}$, which yield the same 
background charge $Q_{R}=b+1/b$.
They correspond to two exponential operators with conformal weight one.
From \eqref{rsd} we then get two possible Schwinger-Dyson equations
\begin{eqnarray}\label{1}
&& \!\!\!\!\!\! \!\!\!\!\!\! \!\!\!\!\!\!\!\!\!\!\!\! \!\!\!\!\!\!\!\!\!\!\!\
[\tilde{\alpha}-Q_R]\langle\prod_{i=1}^{n}e^{2\alpha_{i}\phi(x_{i})}\rangle=\mu_1\, b \int d^{2}x \sqrt{ g(x)}\langle \prod_{i=1}^{n}e^{2\alpha_{i}\phi(x_{i})}e^{2b\phi(x)}\rangle~,\\ \label{2}
&& \!\!\!\!\!\! \!\!\!\!\!\! \!\!\!\!\!\!\!\!\!\!\!\! \!\!\!\!\!\!\!\!\!\!\!\
[\tilde{\alpha}-Q_R]\langle\prod_{i=1}^{n}e^{2\alpha_{i}\phi(x_{i})}\rangle={\mu}_2 \frac{1}{b} \int d^{2}x \sqrt{g(x)}\langle \prod_{i=1}^{n}e^{2\alpha_{i}\phi(x_{i})}e^{\frac{2}{b}\phi(x)}\rangle~,
\end{eqnarray}
where the first coincides with \eqref{integral relation} and the second with \eqref{sc2} at $\mu_1=\mu$ and $\mu_2=\tilde\mu$, respectively. These cosmological constants are renormalized according to the choice of $\beta$.

To analyze the scaling properties of the cosmological constant we consider the exponential 
interaction $\mu_{B}e^{2\alpha\phi(x)}$,
which is regularized by $$\mu_{B}e^{(2\alpha\phi(x)-2\alpha^{2}G(x,x))},$$ where $G(x,x)$ is the two point
Green's function at coincide points. Introducing a regulator $\Lambda$ for $G(x,x)$ by $$G(x,x)\sim\frac{1}{2}{\rm ln}[\frac{\Lambda^{2}}{\mu_{B}}],$$
and performing the redefinition $\mu_{B}\to\Lambda^{2}\mu_{0}$, 
with a dimensionless parameter $\mu_{0}$, the regularized exponent can be written as $\mu_{0}^{1+\alpha^{2}}\Lambda^{2}e^{2\a\phi(x)}$.
Then,  for $\alpha=b$ the dimensionless scaling parameter is
$\mu_{0}^{1+b^{2}}\sim\mu$, while for $\alpha=1/b$ the scaling is $\mu_{0}^{1+1/b^{2}}\sim\tilde{\mu}$.
From these relations it follows that $\tilde{\mu}\sim \mu^{1/b^{2}}$, which is consistent with \eqref{dual constant}.

According to Seiberg \cite{Seiberg} a vertex operator $e^{2\a\phi}$ with $2\mbox{Re}\,\a>Q$ 
can not be defined in Liouville field theory.
Hence, for a given $b$ either $e^{2b\phi(x)}$ or $e^{\frac{2}{b}\phi(x)}$ 
will break the Seiberg bound.
For example, if  $0<b<1$ the Seiberg bound is broken by $e^{\frac{2}{b}\phi(x)}$ and
if $b>1$, then by $e^{{2}{b}\phi(x)}$. For $b=1$ both screening charge operators, as well as, 
equations \eqref{1} and \eqref{2} coincide.
Thus, one has only one screening charge operator for a given $b$.

It is important to note that the Schwinger-Dyson equation is obtained from the Schwinger action
principle \cite{Fujikawa} and the path integral
is the solution of this functional equation.
If one considers the bare Schwinger-Dyson equation as a fundamental equation for Liouville field theory,
then its regularization provides two equations \eqref{1} and \eqref{2}. 
Although the left hand side
of these equations  is the same correlation function, 
the two screening charge operators on the right hand side 
differ and only one of them satisfies
the Seiberg bound for a given value of $b$.
Nevertheless, the equations are valid simultaneously if they are understood
as functional relationships between the analytically continued expectation values. 
These relationships hints towards the use of the Barnes double Gamma function
for the structure constants.

In a follow up to this work we intend to investigate the discussed points from the perspective
proposed in \cite{schnitger}. Namely, to use the operator equation of motion for
the Liouville field in the presence of two screening charges and to show
how these charges define the dual pole structure.

We hope that the Schwinger-Dyson approach discussed in this paper can be extended beyond Liouville field
theory, in particular, to Toda theory and other coset WZW models.

\subsection*{Acknowledgement}
The author thanks George Jorjadze for a collaboration at the initial stage of the work.
He would also like to thank Harald Dorn, Stefan Fredenhagen and Hermann Nicolai
for useful discussions and  comments.
This work has been funded by Erasmus Mundus Joint Doctorate Program by the Grant Number 2010-1816
from the EACEA of the European Commission and also by the IMPRS fellowship.

\end{document}